\documentclass[journal]{IEEEtran}
\usepackage{amsmath,amssymb,amsthm,lipsum} 
\usepackage{cite}
\usepackage{algorithmic}
\usepackage{algorithm}
\usepackage{graphicx}
\usepackage{textcomp}
\usepackage{stfloats}
\usepackage{booktabs}
\usepackage{bm}
\usepackage{color}
\usepackage{subfigure}
\usepackage[T1]{fontenc}

\ifCLASSINFOpdf
\else
\fi
\theoremstyle{Theorem}
\newtheorem{theorem1}{Theorem}

\newtheorem{theorem4}[theorem1]{Theorem}

\theoremstyle{remark}
\newtheorem{rmk1}{Remark}

\theoremstyle{Definition}

\theoremstyle{Lemma}
\newtheorem{lemma1}{Lemma}

\newtheorem{lemma3}[lemma1]{Lemma}

\theoremstyle{Corollary}

\begin{document}
%
\title{\LARGE Blind Orthogonal Least Squares based\\
Compressive Spectrum Sensing}
%
%
%

\author{Liyang~Lu,~\IEEEmembership{Student Member,~IEEE,}
        Wenbo~Xu,~\IEEEmembership{Member,~IEEE,}
        Yue~Wang,~\IEEEmembership{Senior Member,~IEEE,}
        Zhi~Tian,~\IEEEmembership{Fellow,~IEEE}

\thanks{L. Lu and W. Xu are with the Key Lab of Universal
Wireless Communications, Ministry of Education, Beijing University of
Posts and Telecommunications. Y. Wang and Z. Tian are with the Department of Electrical and Computer
Engineering, George Mason University, Fairfax, VA.

W. Xu (xuwb@bupt.edu.cn) and Y. Wang (ywang56@gmu.edu) are the corresponding authors.

}}

%
%

\markboth{IEEE TRANSACTIONS ON VEHICULAR TECHNOLOGY}%
{Shell \MakeLowercase{\textit{et al.}}: Bare Demo of IEEEtran.cls for IEEE Journals}
%



\maketitle

\begin{abstract}
As an enabling technique of cognitive radio (CR), compressive spectrum sensing (CSS) based on compressive sensing (CS) can detect the spectrum opportunities from wide frequency bands efficiently and accurately by using sub-Nyquist sampling rate.
However, the sensing performance of most existing CSS excessively relies on the prior information such as spectrum sparsity or noise variance. Thus, a key challenge in practical CSS is how to work effectively even in the absence of such information. In this paper, we propose a blind orthogonal least squares based CSS algorithm (B-OLS-CSS), which functions properly without the requirement of prior information. Specifically, we develop a novel blind stopping rule for the OLS algorithm based on its probabilistic recovery condition. This innovative rule gets rid of the need of the spectrum sparsity or noise information, but only requires the computational-feasible mutual incoherence property of the given measurement matrix. Our theoretical analysis indicates that the signal-to-noise ratio required by the proposed B-OLS-CSS for achieving a certain sensing accuracy is relaxed than that by the benchmark CSS using the OMP algorithm, which is verified by extensive simulation results.
\end{abstract}


%
\IEEEpeerreviewmaketitle

\section{Introduction}
%
%
%
%
\IEEEPARstart{W}{ith} the rapid deployment of intelligent transport systems (ITSs), spectrum scarcity in vehicular communications is becoming the bottleneck, since the available bandwidth turns to be insufficient to satisfy the requirement for high-quality wireless services facing the high traffic levels in vehicular applications \cite{64}.
	To handle such a challenge, cognitive radio (CR) emerges as a key technology by searching for the unused spectrum resources and providing dynamic spectrum access for secondary users (SUs). To detect as much spectrum opportunities as possible from wide frequency bands at sub-Nyquist sampling rate, compressive spectrum sensing (CSS) methods have been developed based on compressive sensing (CS) and well acknowledged as a promising wideband spectrum sensing solution.
Among various CSS algorithms, the greedy search methods, e.g., orthogonal matching pursuit (OMP) \cite{4} and orthogonal least squares (OLS) \cite{56}, exhibit satisfactory sensing performance with fast implementation. The iterative atom selection mechanism of greedy algorithms, however, relies on the spectrum sparsity or noise prior information, which is not always available in practice and thus hinders their applications.

Blind greedy (BG) algorithms have been developed to solve the aforementioned dilemma of requiring prior information \cite{55}. In current literature, the blind OMP (B-OMP) algorithm, as a representative BG algorithm, keeps detecting the effective support atomic energy in the residuals blindly \cite{4}. However, the performance of OMP is sensitive to the mutual incoherence property (MIP) \cite{60,2} of the measurement matrix, that is, MIP should be small enough for effective atom separation, which limits the applicability of B-OMP algorithms in practice.

By contrary, the OLS algorithm enjoys stronger capability for correct atom exploration than OMP, resulting in compelling spectrum recovery performance, even if the measurement matrix exhibits unsatisfactory MIP \cite{9}. Therefore, OLS is capable to guarantee more stable spectrum access of SUs when different measurement matrices are used in practice, which motivates us to investigate blind OLS algorithm for reliable CSS performance without prior information. To the best of our knowledge, there is no study on developing blind stopping rule for OLS. Accordingly, there is no OLS-related blind algorithm design and performance analyses in the current literature of both CS and CSS based CR.

To fill such a technical gap, this paper proposes a blind OLS-based CSS (B-OLS-CSS) algorithm for CR. Specifically, we formulate the bounds of a mapping factor in OLS, which is tighter than the existing ones, by utilizing the probabilistic norm bound and computational-friendly MIP metric. Then, a blind stopping rule for the OLS algorithm is developed via utilizing the MIP-based recovery conditions. To protect primary users' (PUs') uninterrupted communications and facilitate SUs' spectrum access, our stopping rule focuses on the selection of all correct support atoms. Our work also theoretically demonstrates that the signal-to-noise ratio (SNR) required for reliable recovery of B-OLS-CSS is lower than that required by the blind OMP-based CSS \cite{4}.

The rest of this paper is organized as follows. In Section II, we introduce notations and system model. In Section III, we present our proposed blind stopping rule, B-OLS-CSS algorithm, and the theoretical analysis. In Section IV, simulation results are given, followed by
conclusions in Section V.	

\section{Preliminaries}
\subsection{Notations}
$\mathbf{D}_{\mathbf{S}^l}$ is a submatrix of $\mathbf{D}$ that contains the column set $\mathbf{S}^l$ selected at the $l$-th iteration.
$\mathbf{P}_{\mathbf{S}^l}=\mathbf{D}_{\mathbf{S}^l}\mathbf{D}_{\mathbf{S}^l}^\dag$ denotes the projection onto the ${\rm span}(\mathbf{D}_{\mathbf{S}^l})$, where $\mathbf{D}_{\mathbf{S}^l}^\dag$ represents the pseudoinverse of $\mathbf{D}_{\mathbf{S}^l}$. $\mathbf{P}_{\mathbf{S}^l}^\bot=\mathbf{I}-\mathbf{P}_{\mathbf{S}^l}$ 
represents the projection onto the orthogonal complement of the ${\rm span}(\mathbf{D}_{\mathbf{S}^l})$. 
The spectral norm of a matrix $\mathbf{D}$ is denoted by $\rho(\mathbf{D})$. The measurement matrix is normalized throughout the paper.

\subsection{System Model}

In CR, the received spectrum at a SU is denoted by $\mathbf{s}\in\mathcal{R}^{N}$, which is sparse based on a certain basis $\mathbf{\Psi}\in \mathcal{R}^{N\times N}$. Let $\mathbf{s}=\mathbf{\Psi} \mathbf{x}$, where $\mathbf{x}$ is a $K$-sparse spectrum that only contains $K$ nonzero spectrum support entries.
Define $\mathbf{\Phi}\in \mathcal{R}^{M\times N}$ as the sampling matrix, where $M$ and $N$ are the numbers of sub-Nquist-rate and Nquist-rate samples, respectively. Denoting the additive noise as $\mathbf{\epsilon}\sim \mathcal{N}(\mathbf{0},\sigma^2\mathbf{I}_M)$, the compressed measurement vector $\mathbf{y}\in \mathcal{R}^{M}$ is given by \begin{equation}\label{measurementy}
\mathbf{y}=\mathbf{\Phi} \mathbf{s}+\mathbf{\epsilon}=\mathbf{\Phi}\mathbf{\Psi} \mathbf{x}+\mathbf{\epsilon}=\mathbf{D}\mathbf{x}+\mathbf{\epsilon},
\end{equation}
where $\mathbf{D}=\mathbf{\Phi}\mathbf{\Psi}\in \mathcal{R}^{M\times N}$ is the measurement matrix.
Define ${\rm SNR}=\frac{\mathbb{E}(||\mathbf{D}\mathbf{x}||_2^2)}{\mathbb{E}(||\mathbf{\epsilon}||_2^2)}$, ${\rm SNR}_q=\frac{||\mathbf{x}_q\mathbf{D}_q||_2^2}{M\sigma^2}$ 
and
${\rm SNR}_{\min}$ as the minimum value of ${\rm SNR}_q$ $(q=1,2,\cdots,N)$ \cite{4}.

\begin{figure}
  \centering
  \includegraphics[scale=0.25]{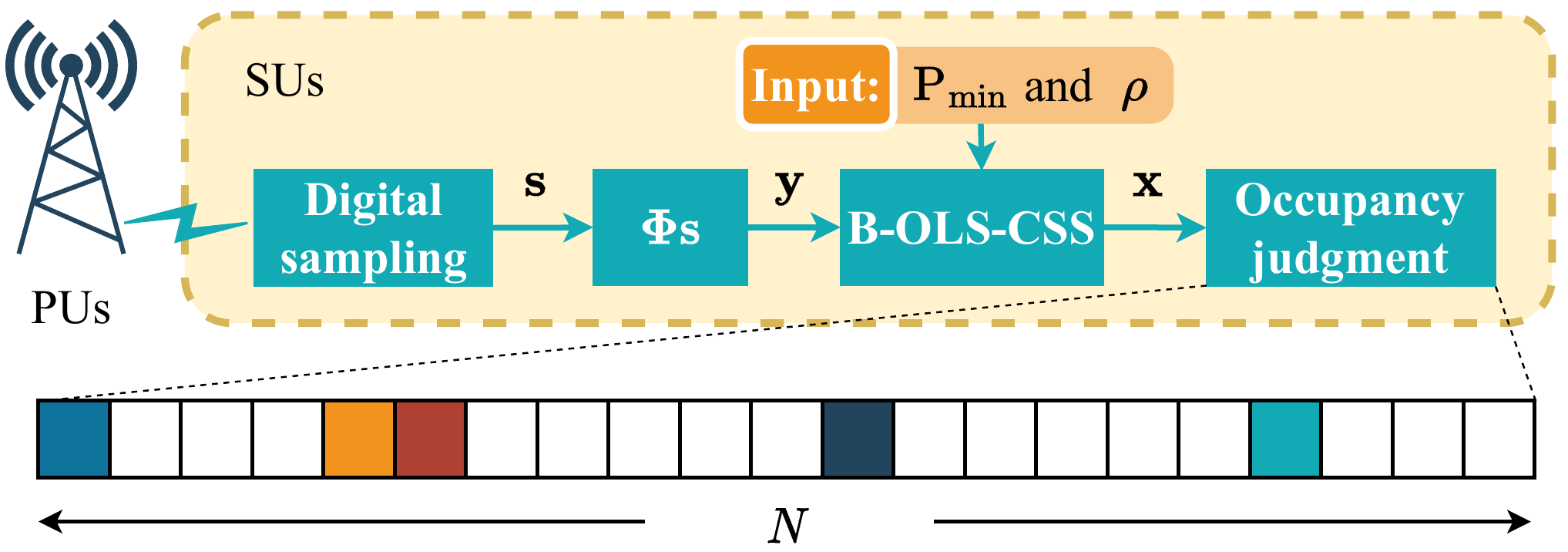}\\
  \caption{System model with B-OLS-CSS algorithm.}\label{systemmodel}
\end{figure}

The objective of CSS is to recover the sparse spectrum $\mathbf{x}$ from the compressed measurement signal $\mathbf{y}$ given the measurement matrix $\mathbf{D}$. The system model with our proposed B-OLS-CSS algorithm is illustrated in Fig. \ref{systemmodel}. At a SU's node, the original sparse spectrum is recovered by our proposed B-OLS-CSS algorithm with the required minimum probability of recovery ${\rm P}_{\min}$ and the parameter $\rho$ that guarantees more opportunities of selecting correct atoms for the algorithm. ${\rm P}_{\min}$ and $\rho$ will be elaborated afterwards.

\section{Proposed compressive spectrum sensing algorithm}

In practical vehicular communications, the sparsity value or the noise information is generally not known \emph{a priori}. Therefore, the conventional stopping rule of OLS is not applicable any more, which indicates that OLS would stop iteration if the preset number of iterations is larger than the ground truth sparsity value $K$ or if the residual is smaller than a predefined threshold according to the noise variance. 
	In this section, to overcome these issues, we first develop a blind stopping rule for OLS based on our analytical results of sparse recovery condition which is independent to the signal sparsity value and the noise statistical information. Then, the B-OLS-CSS algorithm is developed based on the design of the blind stopping rule. 

\subsection{Blind Stopping Rule and Theoretical Analysis for OLS}

According to the iterative procedures of OLS, the residual vector after $l$ iterations $(1\leq l\leq K)$ is
$\mathbf{r}^l=\mathbf{P}_{\mathbf{S}^l}^{\bot}(\mathbf{D}\mathbf{x}+\mathbf{\epsilon})$.
As proved in \cite{4}, $\frac{||\mathbf{D}^T\mathbf{r}^l||_{\infty}}{||\mathbf{r}^l||_2}$ can be used to detect whether there remains nonzero components in the residual vector. That is, if $\frac{||\mathbf{D}^T\mathbf{r}^l||_{\infty}}{||\mathbf{r}^l||_2}$ is smaller than a predefined threshold $\mathcal{Q}$, i.e.,
\begin{equation}\label{conventionalthreshold}
	\frac{||\mathbf{D}^T\mathbf{r}^l||_{\infty}}{||\mathbf{r}^l||_2}<\mathcal{Q},
\end{equation}
the residual vector only contains noise components and OLS stops iteration. In the following, we develop this critical threshold $\mathcal{Q}$ that needs to be predefined by utilizing the analytical results for OLS.
 In doing so, we first present some useful lemmas.
In OLS, the tightness of the mapping factor, i.e., $||\mathbf{P}^{\bot}_{\mathbf{S}^l}\mathbf{D}_i||_2$ $(i\in\{1,2,\cdots,N\}\backslash \mathbf{S}^l)$, determines the tightness of theoretical recovery condition \cite{65}. The following lemmas present a tighter bound for the mapping factor than the existing results \cite{1,2}.

\begin{lemma1}
	For $\mathbf{B}\in \mathcal{R}^{M\times K}$, whose entries independently and identically satisfy  $\mathcal{N}(0,\frac{1}{M})$, the smallest singular value $\zeta_{\min}$ and the largest singular value $\zeta_{\max}$ with any $\rho>0$ follow:
\begin{equation}\label{largestsandsmallest1}
	\begin{aligned}
		\min\bigg\{&{\rm P}\Big\{\zeta_{\min}\geq1-\sqrt{K/M}-\rho\Big\},\\
		&{\rm P}\Big\{\zeta_{\max}\leq1+\sqrt{K/M}+\rho\Big\}\bigg\}
		\geq1-e^{-\frac{M\rho^2}{2}}.
	\end{aligned}
\end{equation}
\label{lemma1}
\end{lemma1}

The proof of Lemma 1 is omitted since it can be easily derived from \cite[Theorem 2.13]{57}.

\begin{lemma3}
	Suppose $\mu<\frac{1}{K-1}$, then $\frac{1}{\sqrt{\mathcal{T}}}\leq||\mathbf{P}^{\bot}_{\mathbf{S}^l}\mathbf{D}_i||_2\leq1$ for $i\in\{1,2,\cdots,N\}\backslash \mathbf{S}^l$ with the probability given in Lemma~\ref{lemma1},
where $\mathcal{T}=\Big(1-\frac{K\mu^2(1+\sqrt{K/m}+\rho)}{(1-\sqrt{K/m}-\rho)^2}\Big)^{-1}$.
\label{lemma3}
\end{lemma3}
\begin{IEEEproof}
See Appendix \ref{proofoflemma3}.
\end{IEEEproof}

The closer the lower bound in Lemma \ref{lemma3} is to 1, the tighter it is.
The tightness of this bound depends on the parameter $\rho$. Next, compared with two existing bounds of the mapping factor, we discuss the bound in our derived Lemma \ref{lemma3} in terms of the range of $\rho$ which is tighter than the existing ones. In \cite{1} and \cite{2}, the authors provide that
\begin{equation}\label{existingone}
	\sqrt{1-K\mu}\leq||\mathbf{P}_{\mathbf{S}^l}^{\bot}\mathbf{D}_i||_2\leq1
\end{equation}
and
\begin{equation}\label{existingtwo}
	\sqrt{1-\frac{(1+(K-1)\mu)K\mu^2}{(1-(K-1)\mu)^2}}\leq||\mathbf{P}_{\mathbf{S}^l}^{\bot}\mathbf{D}_i||_2\leq1,
\end{equation}
respectively. The comparison between our result in Lemma 2 and those in (\ref{existingone}) and (\ref{existingtwo}) is presented in the following remark.

\begin{rmk1}
If $\rho<(K-1)\mu-\sqrt{K/M}$, $(||\mathbf{P}_{\mathbf{S}^l}^{\bot}\mathbf{D}_i||_2)$'s lower bound in Lemma~\ref{lemma3} is closer to 1 than those in (\ref{existingone}) and (\ref{existingtwo}).
\label{rmk1}
\end{rmk1}

Remark \ref{rmk1} reveals that the bound of the mapping factor is tighter than the existing results \cite{1,2}. To derive the recovery condition for standard OLS, we utilized the upper bound of the reconstructible sparsity, which is the Theorem 2 in our previous work \cite{2}. It indicates that if the real sparsity $K$ of the spectrum is lower than a specific threshold $\mathcal{C}$, OLS produces reliable recovery. Note that the threshold $\mathcal{C}$ is only related with the matrix coherence of the measurement matrix.
Based on the theoretical analysis related to the mapping factor and the reconstructible sparsity, we present the reliable recovery condition for OLS.

\begin{theorem4}
Suppose $K<\mathcal{C}$ and $\mathbf{\epsilon}\sim\mathcal{N}(0,\sigma^2\mathbf{I}_M)$. Define $\mathcal{P}(\omega)
=1-\frac{\mathcal{C}N}{e^{0.5\omega^2\mu^2\theta^2}\sqrt{2\pi\omega^2\mu^2\theta^2}}-2e^{-\frac{M\rho^2}{2}}
-\frac{1}{M-\mathcal{C}}-\frac{1}{M}$,
where
$\theta=\sqrt{\mathcal{A}_1}-\sqrt{\mathcal{A}_2}$,
$\mathcal{A}_1=4(M-\mathcal{C})-2$, $\mathcal{A}_2=M-\mathcal{C}+2\sqrt{(M-\mathcal{C})\log(M-\mathcal{C})}$ and $\rho>0$.

For a given minimum probability of recovery $\mathrm{P}_{\min}$, the OLS algorithm using the stopping rule (3), i.e.,  $\frac{||\mathbf{D}^T\mathbf{r}^l||_{\infty}}{||\mathbf{r}^l||_2}<\mathcal{Q}$,
can reconstruct the $K$-sparse spectrum with the probability $\mathrm{P}>\mathrm{P}_{\min}$,
if the minimum component $\mathrm{SNR}_{\min}$ satisfies
\begin{equation}\label{theoremmainSNRrelation}
	\begin{aligned}
		\mathrm{SNR}_{\min}>\max\{\varphi_1,\varphi_2\},
	\end{aligned}
\end{equation}
where $\varphi_1=\frac{4(2-(K-\mathcal{T})\mu)^2\omega^2\mu^2\theta^2}{M(2-(K-\mathcal{T})\mu-2K\mathcal{T}\mu)^2(1-(K-1)\mu)^2}$, $\varphi_2=\frac{\omega^2\mu^2(\theta+\sqrt{M+2\sqrt{M\log M}})^2}{M(1-\sqrt{K/M}-\rho-\omega\mu(1+\sqrt{K/M}+\rho)\sqrt{K})^2}$, $\mathcal{Q}=\omega\mu$, $\omega=\mathcal{P}^{-1}(\mathrm{P}_{\min})$, $\mathcal{P}^{-1}(\cdot)$ represents the inverse function of $\mathcal{P}(\cdot)$.

\label{theorem4}
\end{theorem4}
\begin{IEEEproof}
See Appendix \ref{proofoftheorem4}.
\end{IEEEproof}

It is observed that the right-hand-side of the stopping rule $\mathcal{Q}=\omega\mu$ contains a constant $\omega=\mathcal{P}^{-1}(\mathrm{P}_{\min})$ and the computable matrix coherence $\mu$. It indicates that Theorem~\ref{theorem4} is operational with an input target probability of recovery $\mathrm{P}_{\min}$ and the calculated $\mu$. That is, the stopping rule in Theorem \ref{theorem4} can work blindly since it is independent to the sparsity level or the noise prior information. Thus, it is suitable for practical CSS scenario where this information is unavailable.

\subsection{B-OLS-CSS Algorithm}
In this subsection, we utilize Theorem \ref{theorem4} to develop the B-OLS-CSS algorithm. The proposed B-OLS-CSS algorithm is given in Algorithm 1. Note that the greedy algorithms may not select the exactly correct $K$ support atoms within $K$ iterations due to the presence of noise. The B-OLS-CSS algorithm is designed to address this problem by appropriately reducing the right-hand-side of the blind stopping rule in Theorem \ref{theorem4} and hence the algorithm runs more than $K$ iterations for more opportunities to choose all the $K$ correct support atoms to the best effort. Based on these arguments, the blind stopping rule in the B-OLS-CSS algorithm is set as $\frac{||\mathbf{D}^T\mathbf{r}^l||_{\infty}}{||\mathbf{r}^l||_2}\leq\omega^*\mu$,
where $\omega^*=\omega-\rho$.
According to Remark \ref{rmk1} and Theorem \ref{theorem4}, $\rho$ satisfies  $0<\rho<(\mathcal{C}-1)\mu-\sqrt{\mathcal{C}/M}$. A large $\rho$ induces a loose bound of $||\mathbf{P}_{\mathbf{S}^l}^{\bot}\mathbf{D}_i||_2$, while a small one reduces the expected probability exponentially.
Hence, a moderate scale of $\rho$ is able to effectively balance the required ${\rm SNR}_{\min}$ and the expected recovery probability in Theorem 1. The trade-off ensures that B-OLS-CSS iterates slightly more than $K$ times.

\begin{algorithm}[t]
	\renewcommand{\algorithmicrequire}{\textbf{Input:}}
	\renewcommand{\algorithmicensure}{\textbf{Output:}}
	\begin{algorithmic}[1]
		\REQUIRE The measurement matrix $\mathbf{D}$, compressed measurements $\mathbf{y}$, minimum target probability of recovery ${\mathrm P}_{\min}$, and parameter $\rho$.
		\ENSURE The recovered spectrum $\mathbf{x}\in\mathcal{R}^{N}$, and recovered index set of nonzero entries $\mathbf{S}\subseteq \{1,2,\cdots,N\}$.
		\STATE $\mathbf{Initialization:}$ $l=0$, $\mathbf{r}^0=\mathbf{y}$, $\mathbf{S}^0=\emptyset$, $\mathbf{x}^0=\mathbf{0}$.
		\STATE Calculate $\omega=\mathcal{P}^{-1}({\mathrm P}_{\min})$ and set $\omega^*=\omega-\rho$, where $\mathcal{P}^{-1}(\cdot)$ is given in Theorem~1.
		\STATE Calculate the matrix coherence $\mu$ of $\mathbf{D}$, where $\mu=\max\limits_{i,j\neq i}|\mathbf{D}_i^{T}\mathbf{D}_j|$.
		\WHILE {$\frac{||\mathbf{D}^T\mathbf{r}_t||_{\infty}}{||\mathbf{r}_t||_2}>\omega^*\mu$}
		\STATE Set $i^{l+1}=\mathop{\arg\min}\limits_{j\in\{1,\cdots,N\}\backslash\mathbf{S}^{l}}||\mathbf{P}^\bot_{\mathbf{S}^{l}\cup \{j\}}\mathbf{y}||_2^2$, \\
		where $\mathbf{P}_{\mathbf{S}^{l}\cup \{j\}}^\bot=\mathbf{D}_{\mathbf{S}^{l}\cup \{j\}}(\mathbf{D}_{\mathbf{S}^{l}\cup \{j\}}^T\mathbf{D}_{\mathbf{S}^{l}\cup \{j\}})^{-1}\mathbf{D}^T_{\mathbf{S}^{l}\cup \{j\}}$;
		\STATE Augment $\mathbf{S}^{l+1}=\mathbf{S}^{l}\cup{\{i^{l+1}\}}$;
		\STATE Estimate $\mathbf{x}^{l+1}=\mathop{\arg\min}\limits_{\mathbf{x}: \;{\mathrm supp}(\mathbf{x})=\mathbf{S}^{l+1}}\|\mathbf{y}-\mathbf{D}\mathbf{x}\|_2^2$;
		\STATE Update $\mathbf{r}^{l+1}=\mathbf{y}-\mathbf{D}\mathbf{x}^{l+1}$;
		\STATE $l=l+1$;
		\ENDWHILE
		\STATE \textbf{return} $\mathbf{S}=\mathbf{S}^{l}$ and $\mathbf{x}=\mathbf{x}^{l}$.
	\end{algorithmic}
	\caption{B-OLS-CSS Algorithm}
	\label{alg:11}
\end{algorithm}

\section{Simulation Results  }

\subsection{Simulations for Theoretical Results}

In this subsection, we perform simulations to compare  Lemma \ref{lemma3} and Theorem \ref{theorem4} with (\ref{existingone}), (\ref{existingtwo}) and \cite[Theorem 1]{4}.

\begin{figure} \centering 
	\subfigure[] { \label{fig:a} 
		\includegraphics[width=0.466\columnwidth]{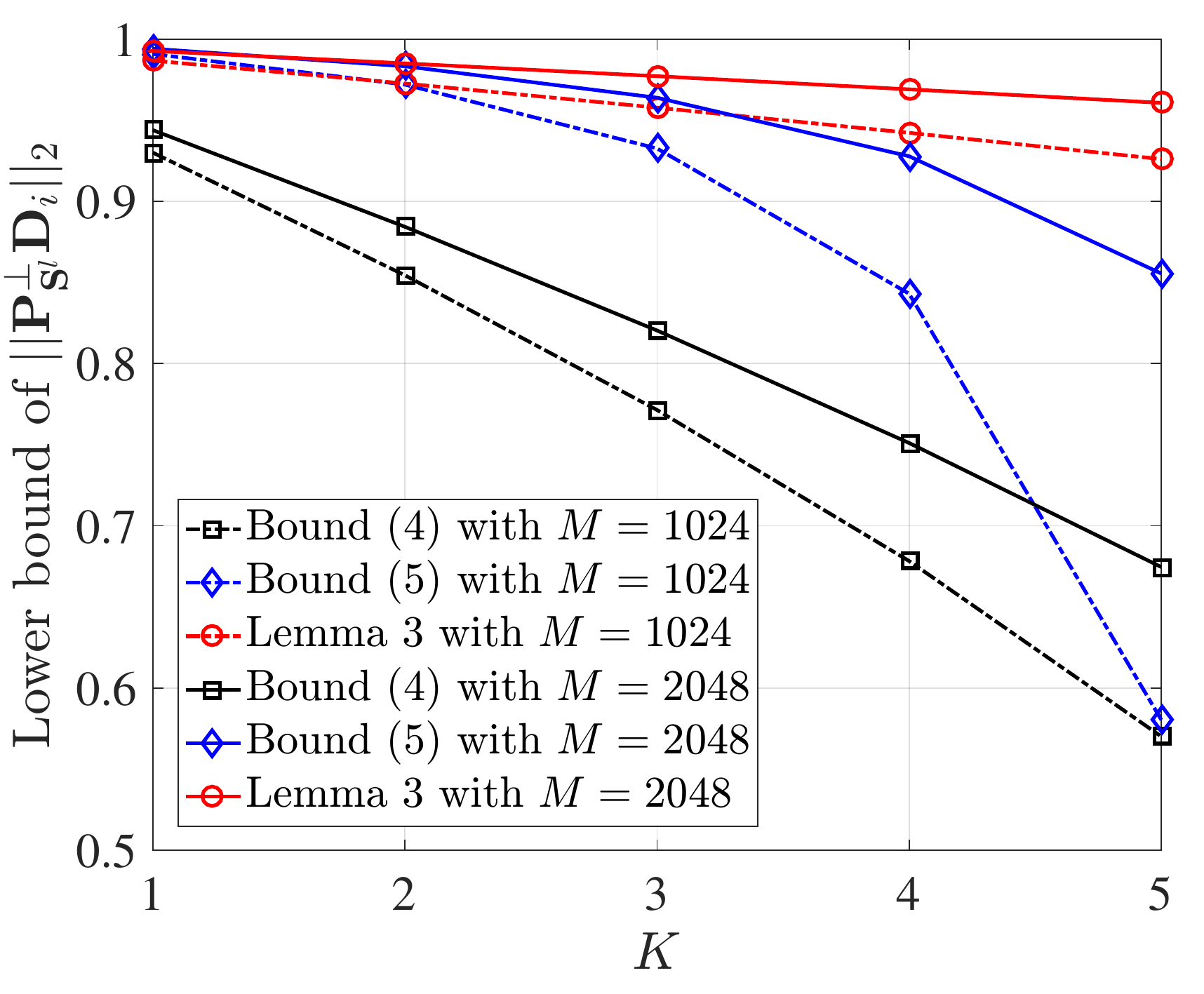} 
	} 
	\subfigure[] { \label{fig:b} 
		\includegraphics[width=0.466\columnwidth]{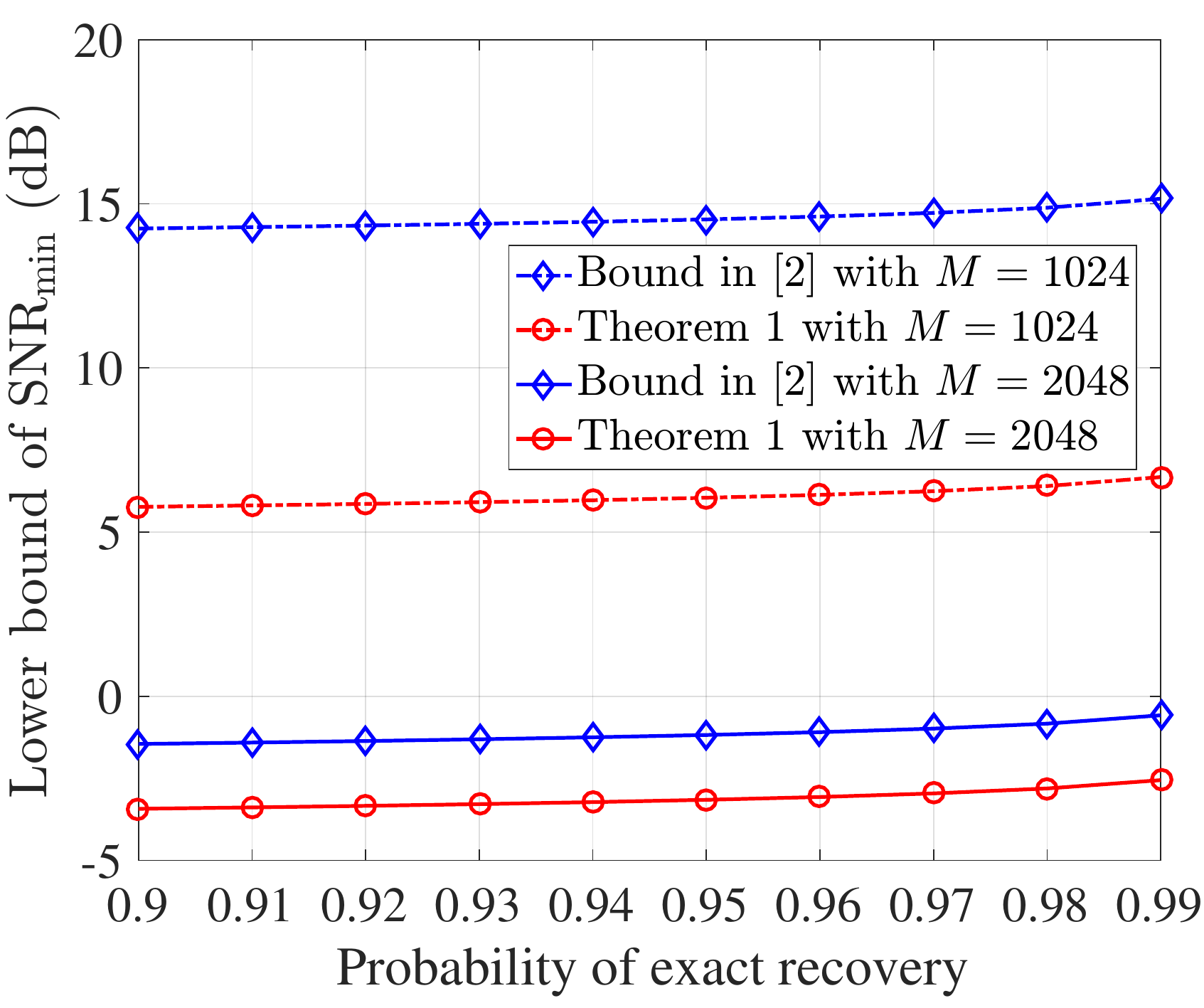} 
	} 
	\caption{Lower bounds of (a) $||\mathbf{P}_{\mathbf{S}^l}^{\bot}\mathbf{D}_i||_2$; (b) ${\rm SNR}_{\min}$ with $K=4$.}
	\label{lowerboundscom} 
\end{figure}

We generate two $M\times N$ normalized measurement matrices (where $M=1024$, $N=8192$ and $M=2048$, $N=8192$), which are the same as those in \cite{4}. The matrix coherences $\mu$ of these matrices are about $0.135$ and $0.109$ respectively. For fair comparison, we fix $\vartheta$ as $0.15$ in the simulations. The results of the comparison among the lower bounds in Lemma~\ref{lemma3}, (\ref{existingone}) and (\ref{existingtwo}) are presented in Fig.~\ref{lowerboundscom}(a). It is observed that our derived bound is generally much closer to 1 than those in (\ref{existingone}) and (\ref{existingtwo}), which verifies that our result is tighter. When matrix coherence $\mu$ is smaller, the lower bound of $||\mathbf{P}_{\mathbf{S}^l}^{\bot}\mathbf{D}_i||_2$ becomes tighter.

The lower bounds of ${\rm SNR_{\min}}$ for high probability of recovery in Theorem \ref{theorem4} and \cite[Theorem 1]{4} are presented in Fig. \ref{lowerboundscom}(b). The recovery probabilities are set to be $0.9:0.01:0.99$. As the matrix coherence decreases, the lower bounds of ${\rm SNR}_{\min}$ is reduced. The results indicate that our derived bound of ${\rm SNR}_{\min}$ for OLS is lower than that of OMP, which implies that OLS is more suitable for CSS in noisy scenarios.

\subsection{Simulations for CSS}

In this subsection, we present simulations to demonstrate the superiority of our proposed B-OLS-CSS algorithm beyond the existing benchmarks. The spectrum is regarded to be successfully recovered if the recovered spectrum is within a certain small Euclidean distance of the ground truth. The locations of the sparse nonzero atoms are selected uniformly at random. The nonzero entries in the spectrum are set to be independently and identically distributed as $\mathcal{N}(1,0.01)$ for illustration. We set ${\rm P}_{\min}=0.95$ and $\rho=0.175$. All CSS algorithms run over $1,000$ Monte Carlo trials.

\begin{figure} \centering 
	\subfigure[] { \label{fig:a} 
		\includegraphics[width=0.465\columnwidth]{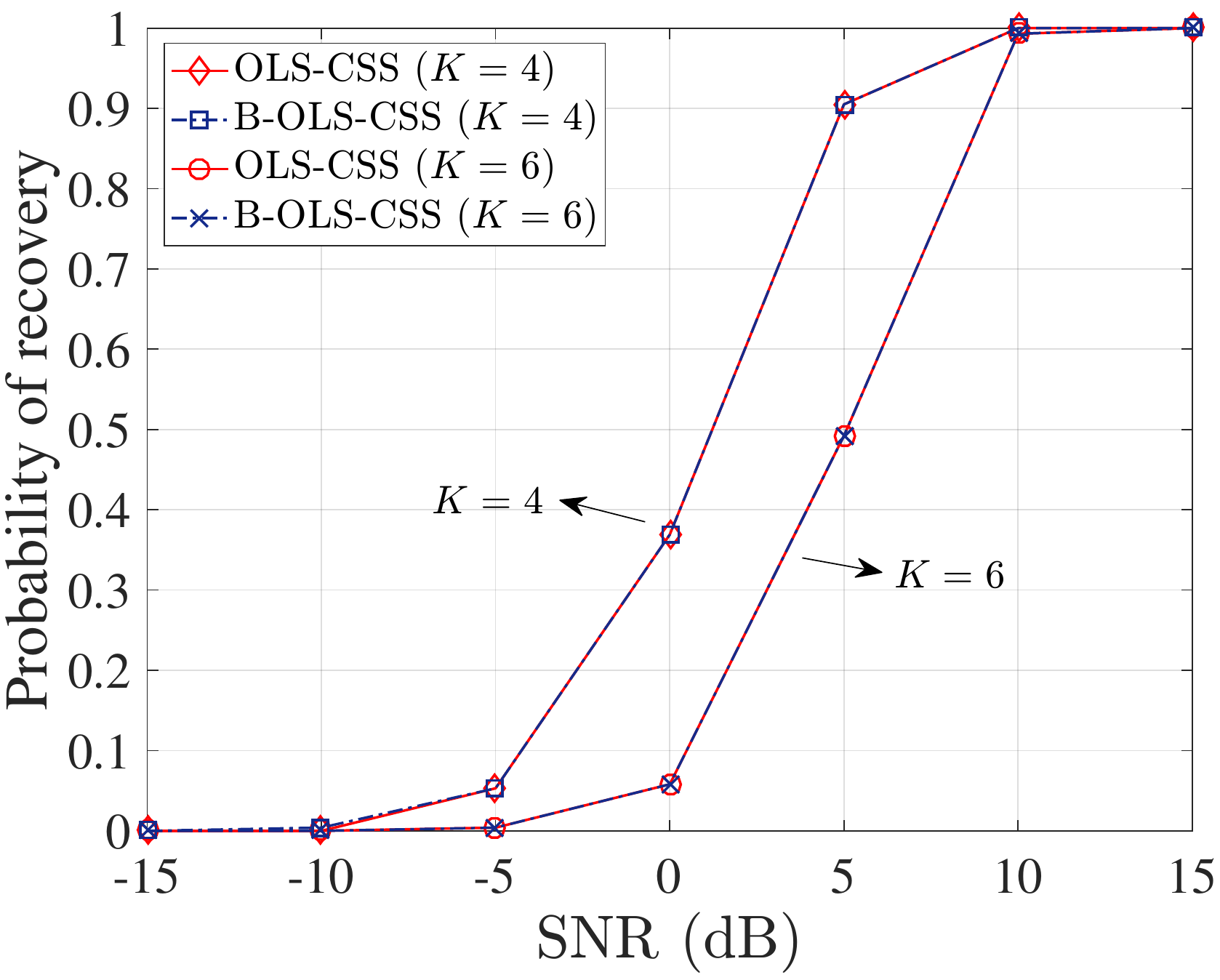} 
	} 
	\subfigure[] { \label{fig:b} 
		\includegraphics[width=0.465\columnwidth]{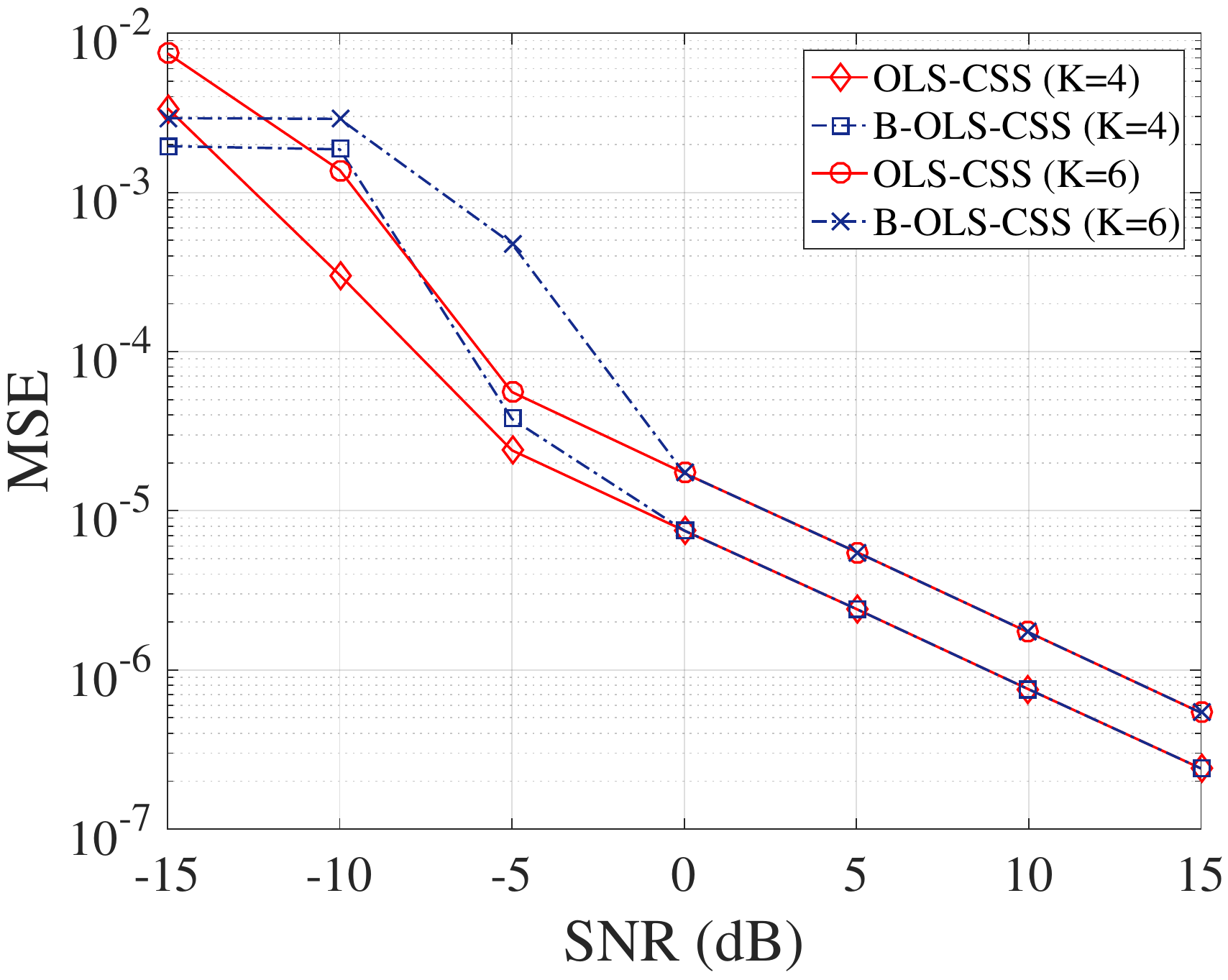} 
	} 
	\caption{(a) Probability of recovery versus SNR (dB) using Gaussian measurement matrix with $M=1024$ and $N=2048$; (b) MSE versus SNR (dB) using Gaussian measurement matrix with $M=1024$ and $N=2048$.} 
	\label{cssolsbolscom} 
\end{figure}

We first perform simulations to compare the sensing performance of the B-OLS-CSS with the OLS-CSS that stops exactly at the $K$-th iteration. As shown in Fig. \ref{cssolsbolscom}, the performance of the B-OLS-CSS is almost the same as that of the OLS-CSS given perfectly known prior knowledge on $K$ and $\sigma$. This shows that even if the sparsity level or the noise information is unavailable, the sensing performance achieved by B-OLS-CSS algorithm is still competitive.

\begin{figure} \centering 
	\subfigure[] { \label{fig:a} 
		\includegraphics[width=0.466\columnwidth]{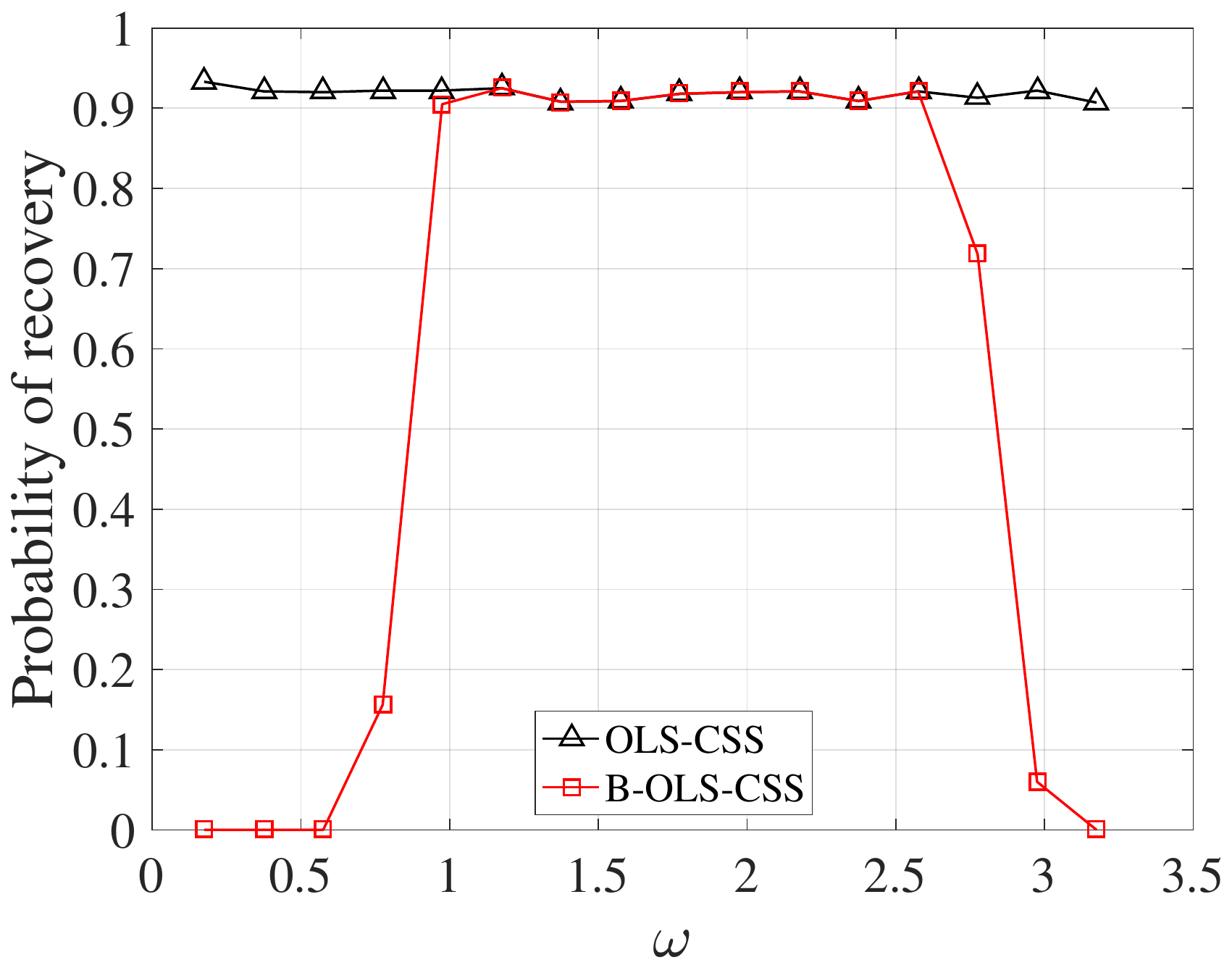} 
	} 
	\subfigure[] { \label{fig:b} 
		\includegraphics[width=0.466\columnwidth]{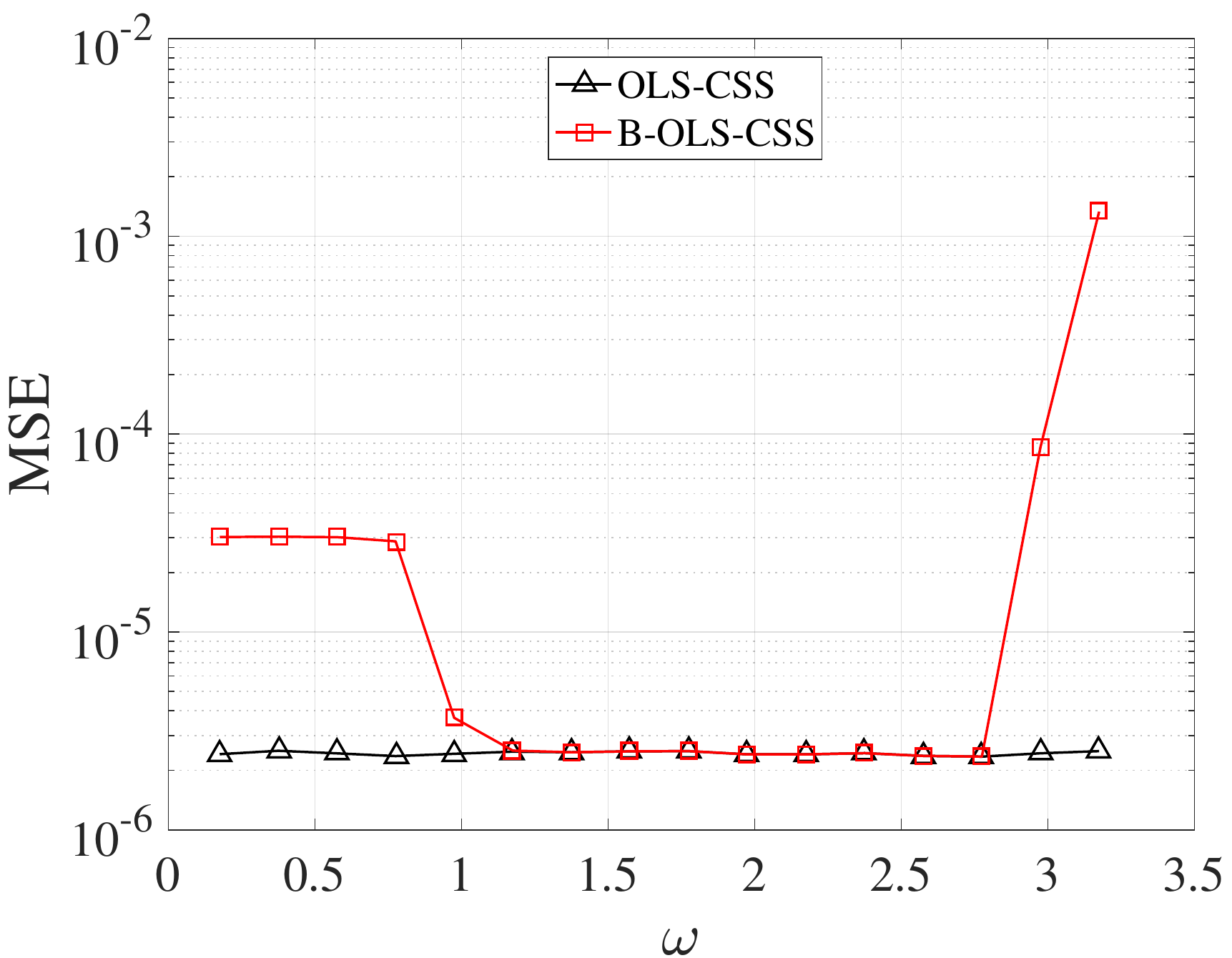} 
	} 
	\caption{(a) Probability of recovery versus $\omega$ using Gaussian measurement matrix with $M=1024$, $N=2048$ and $K=4$; (b) MSE versus $\omega$ using Gaussian measurement matrix with $M=1024$, $N=2048$ and $K=4$.}
	\label{CSSOMEGA} 
\end{figure}

	The sensing performance versus $\omega$ is given in Fig. \ref{CSSOMEGA}. It is observed that the performance of B-OLS-CSS is competitive to that of OLS-CSS in a wide range of $\omega$, which is $[1.175,2.575]$. Such phenomenon indicates that B-OLS-CSS is insensitive to the value of the parameter $\omega$. Based on the settings in Fig. \ref{CSSOMEGA}, the $\omega$ presented in Theorem \ref{theorem4} is approximately equal to 1.3, falling into the range $[1.175,2.575]$. This reveals the guideline role of our derived theoretical results.

Then, we investigate the sensing performance of the OMP-CSS, CoSaMP-CSS, CSS with the blind OMP in \cite{4} (we call it B-OMP-CSS), OLS-CSS, MOLS-CSS \cite{15} and B-OLS-CSS.
 We adopt the hybrid measurement matrix $\mathbf{D}$ given in \cite{9}, whose columns satisfy $\mathbf{D}_i=\mathbf{n}_i+c_i\mathbf{1}$, where $\mathbf{n}_i\sim\mathcal{N}(0,1)$ and $c_i$ obeys the uniform distribution on $[0,10]$. Compared with the Gaussian measurement matrix, the MIP of the hybrid measurement matrix is extremely unsatisfactory. The following simulations indicate that our proposed B-OLS-CSS always has desired performance even for unsatisfactory MIP cases.
	
\begin{figure}
	\centering
	\includegraphics[scale=0.33]{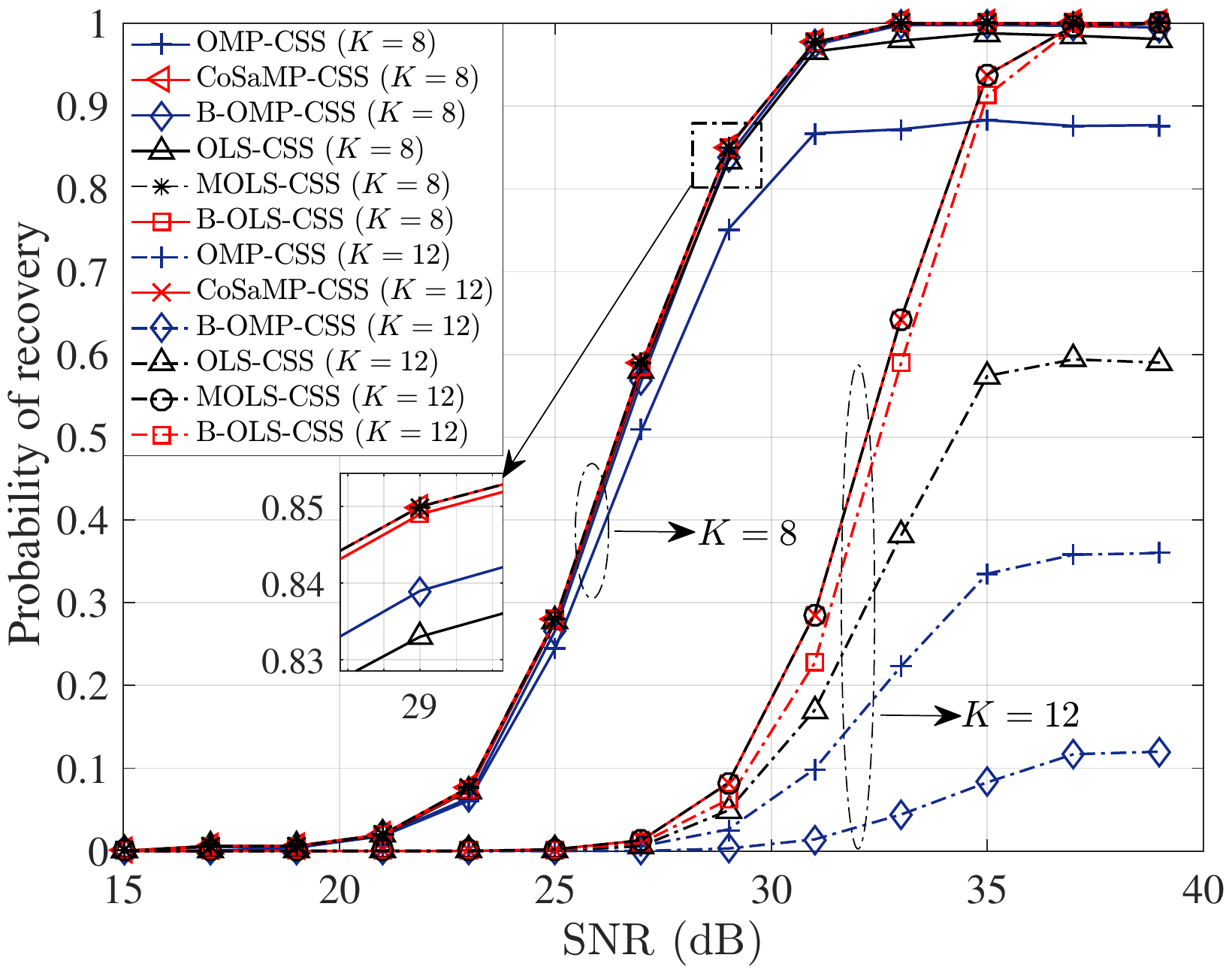}\\
	\caption{Probability of recovery versus SNR (dB) using hybrid measurement matrix with $M=256$ and $N=512$.}\label{css2}
\end{figure}

	As shown in Fig. \ref{css2}, when $K=8$, the performance of B-OLS-CSS is competitive with that of the CoSaMP and MOLS algorithms, and is even better than the OLS-CSS algorithm, which reveals that our proposed blind stopping rule works better than the methods implementing exact $K$ iterations in OLS-CSS. Meanwhile, OMP-CSS behaves the worst, which is consistent with the theoretical statement in \cite{9} that OMP performs poorly in dealing with high coherence measurement matrices. We further conduct the simulations of these algorithms with $K=12$. The performance of OMP-CSS, B-OMP-CSS and OLS-CSS degrades rapidly, and they end up unable to perform reliable recovery. In this case, the performance of our proposed B-OLS-CSS still approaches to that of the CoSaMP and MOLS, which indicates that B-OLS-CSS is more suitable for the practical CR when the prior information is unavailable.

\begin{figure} \centering 
	\subfigure[] { \label{fig:a} 
		\includegraphics[width=0.466\columnwidth]{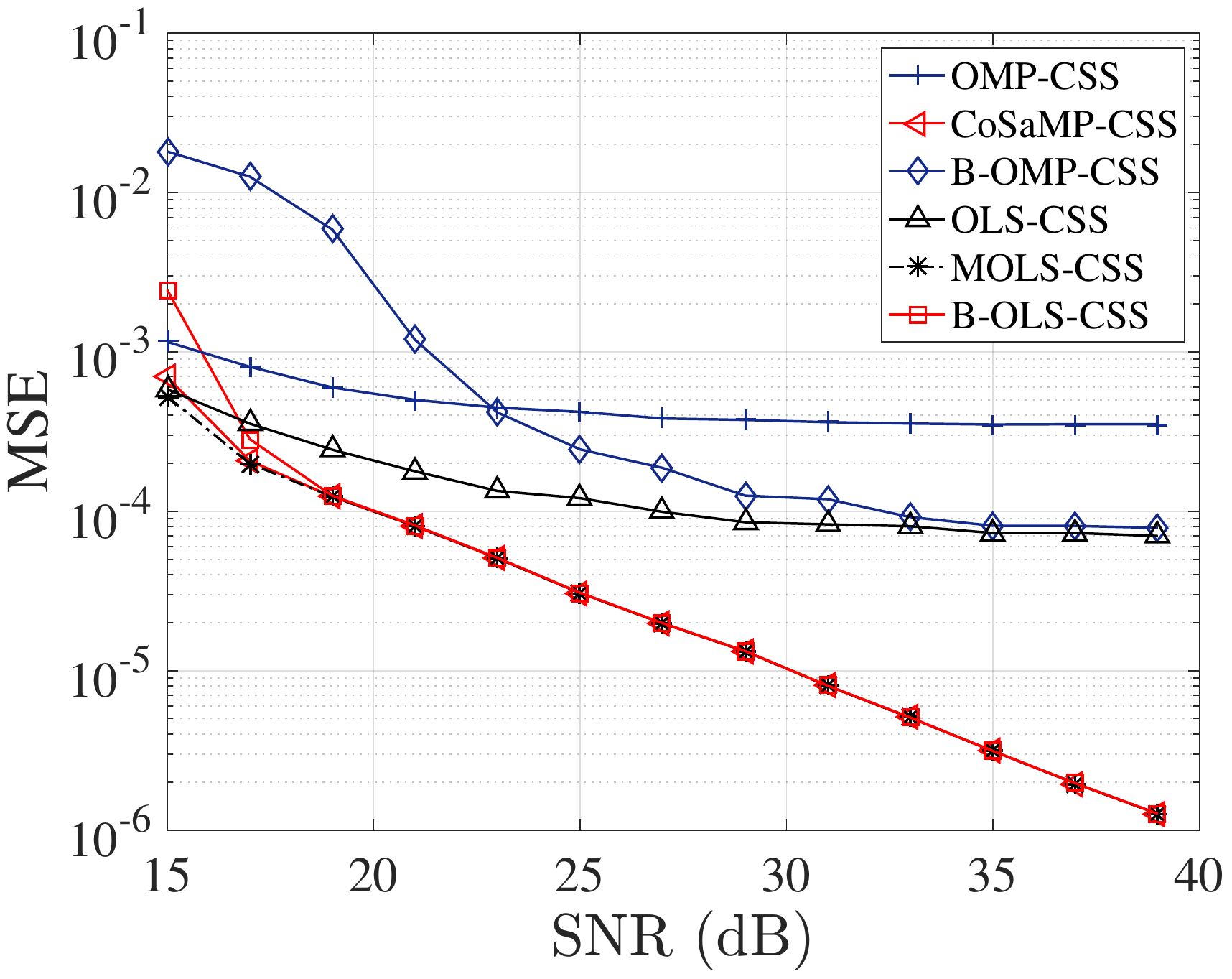} 
	} 
	\subfigure[] { \label{fig:b} 
		\includegraphics[width=0.466\columnwidth]{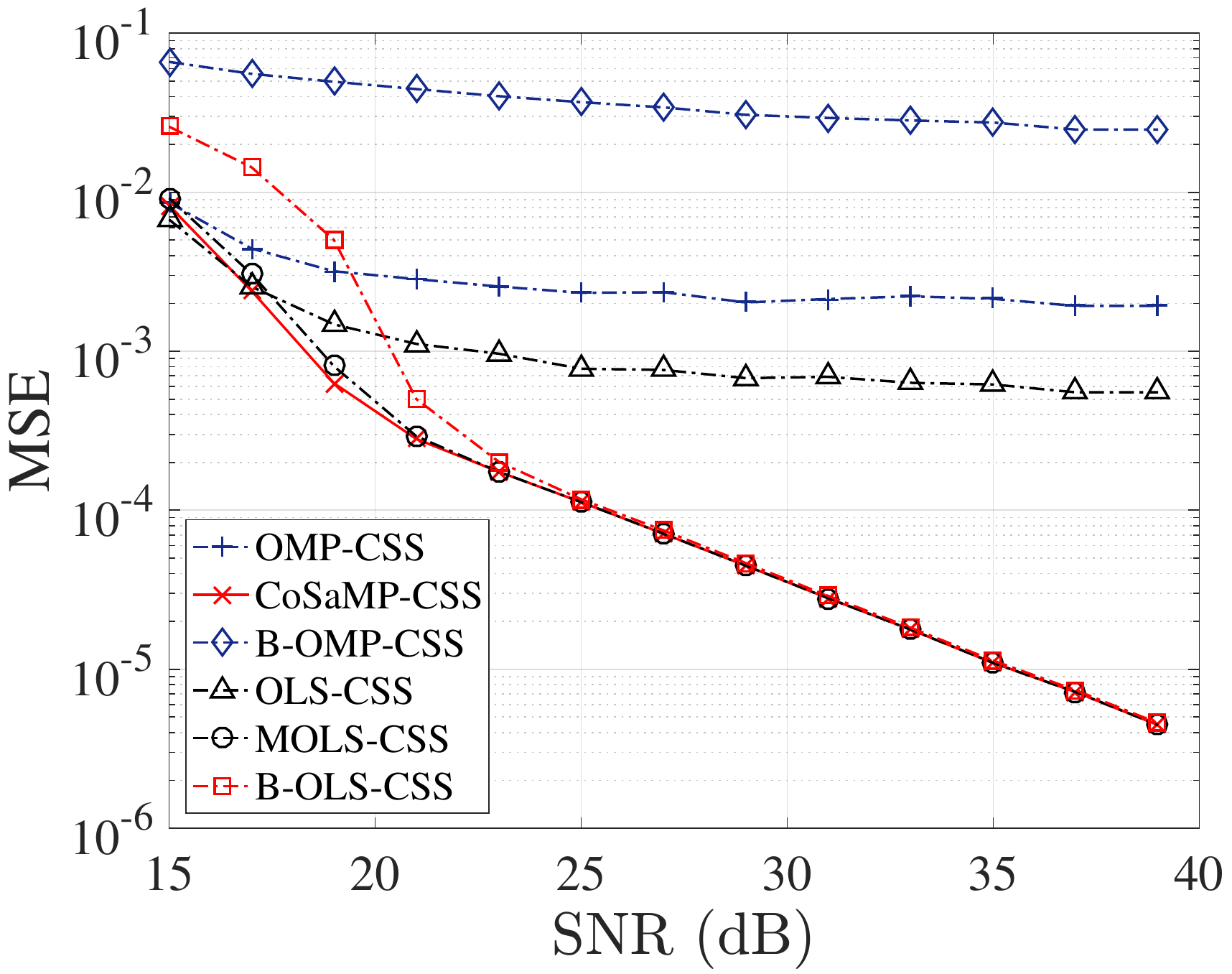} 
	} 
	\caption{MSE versus SNR (dB) using hybrid measurement matrix with $M=256$, $N=512$ and (a) $K=8$; (b) $K=12$.}
	\label{cssmse1} 
\end{figure}

\begin{figure} \centering 
	\subfigure[] { \label{fig:a} 
		\includegraphics[width=0.466\columnwidth]{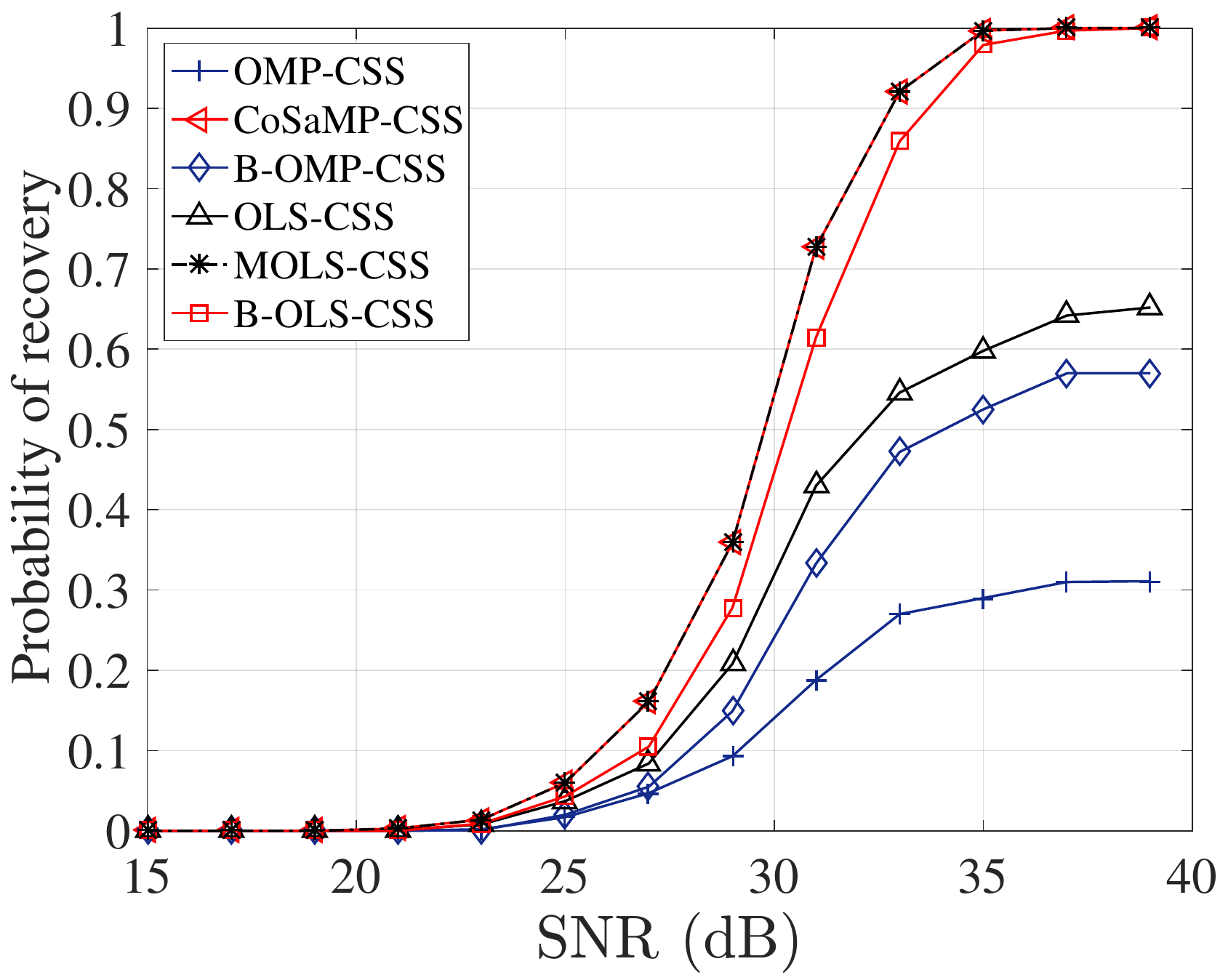} 
	} 
	\subfigure[] { \label{fig:b} 
		\includegraphics[width=0.466\columnwidth]{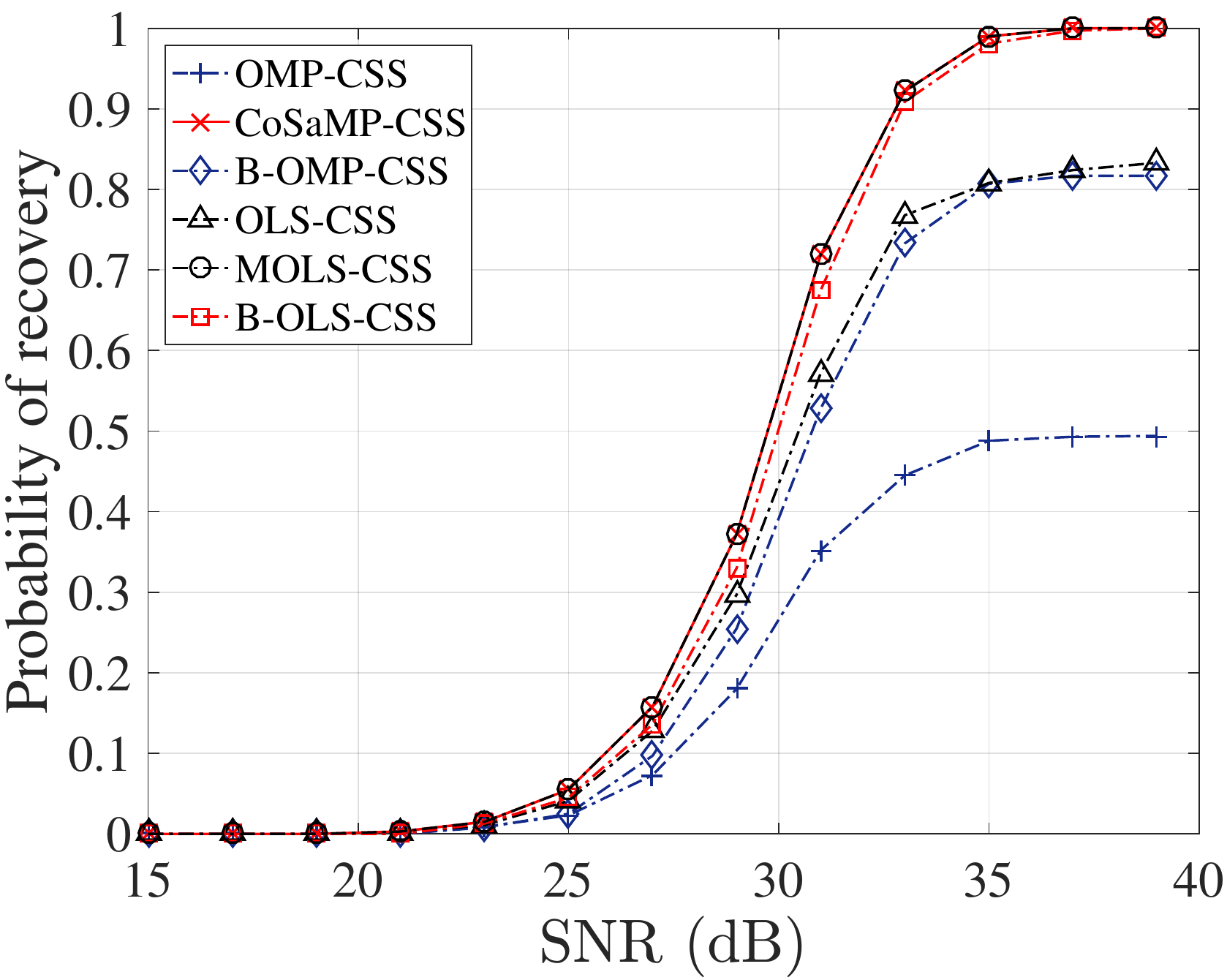} 
	} 
	\caption{Probability of recovery versus SNR (dB) using hybrid measurement matrix with $K=8$ and (a) $M=128$, $N=512$; (b) $M=128$, $N=256$.}
	\label{csspodMNchange2} 
\end{figure}

\begin{figure} \centering 
	\subfigure[] { \label{fig:a} 
		\includegraphics[width=0.466\columnwidth]{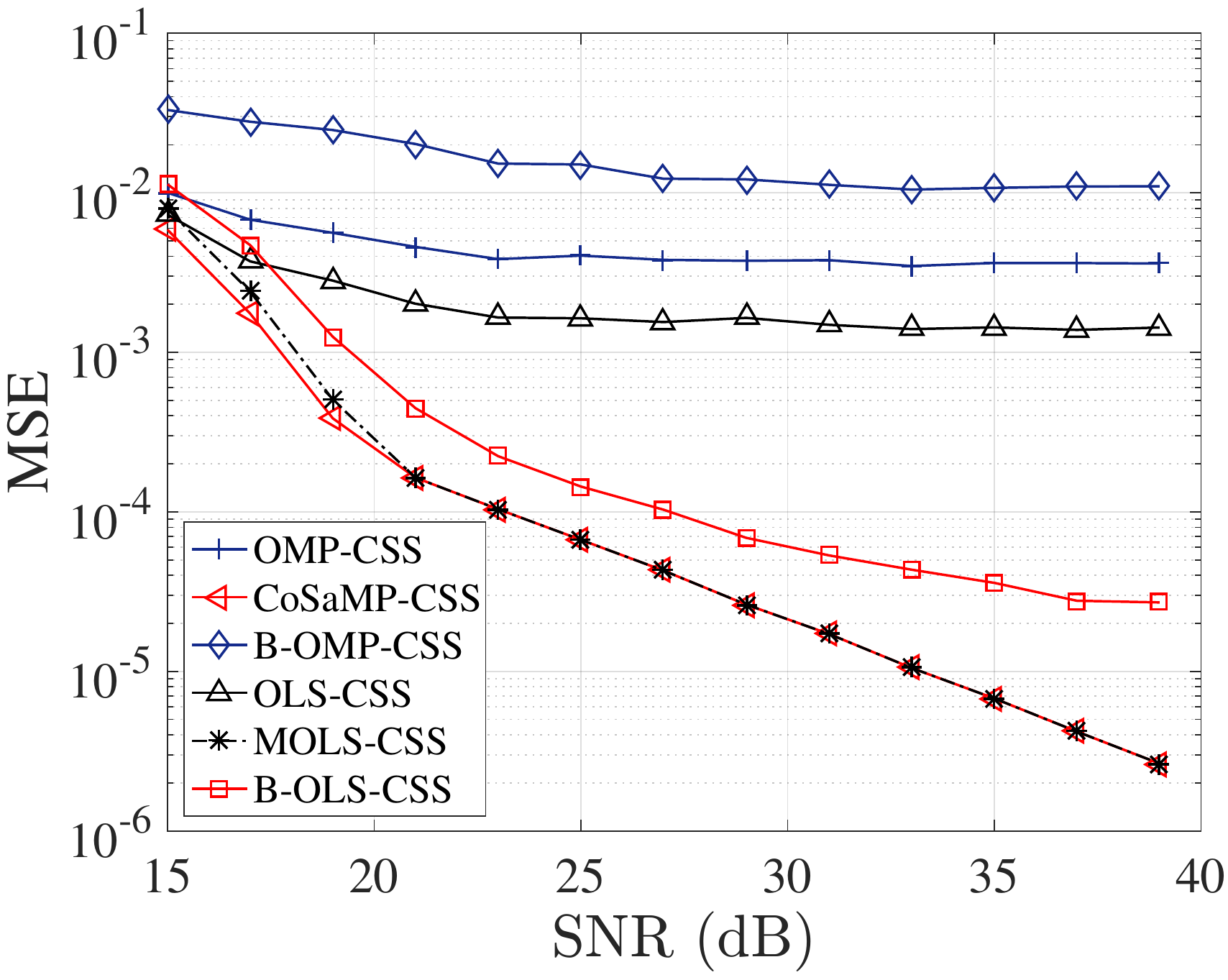} 
	} 
	\subfigure[] { \label{fig:b} 
		\includegraphics[width=0.466\columnwidth]{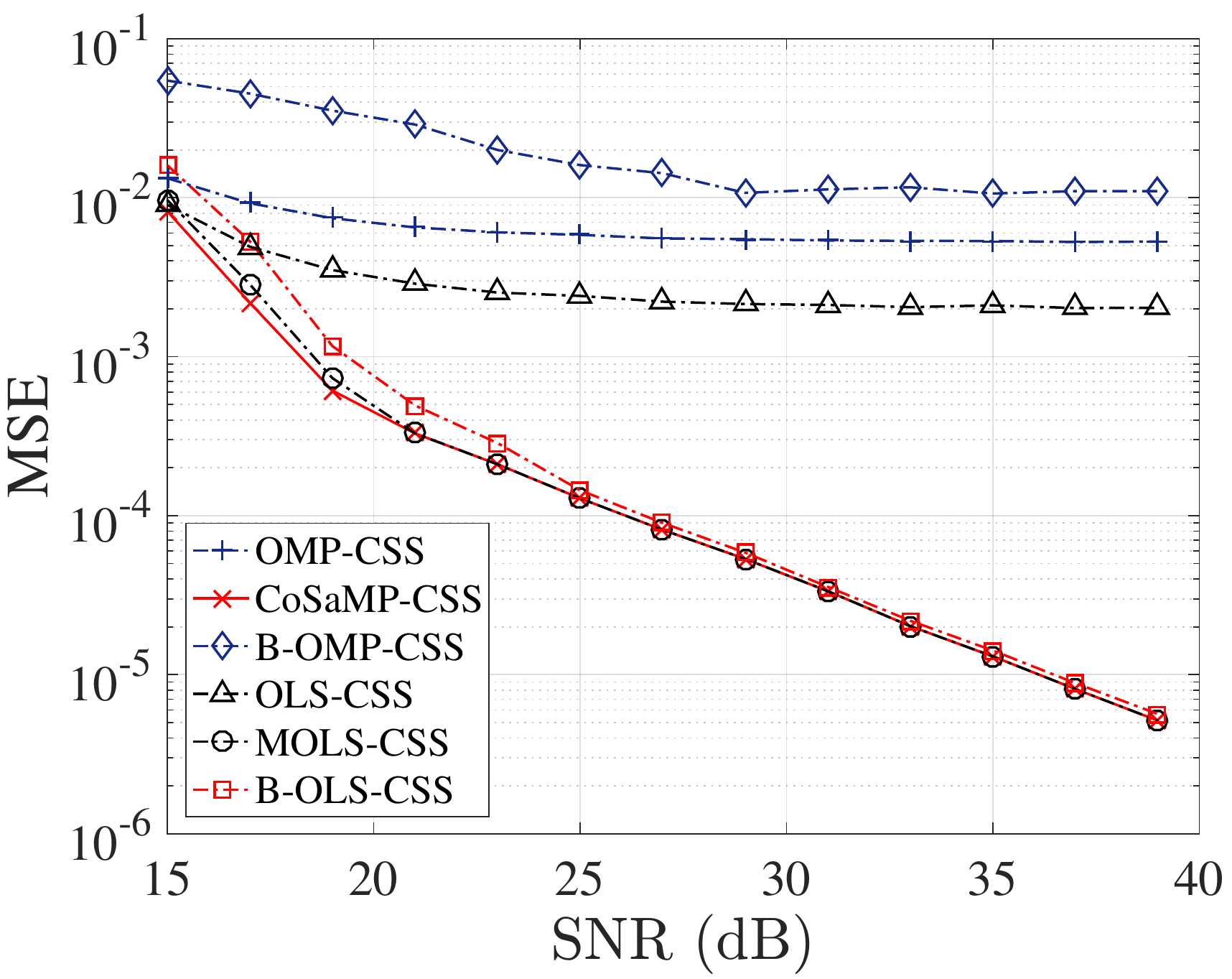} 
	} 
	\caption{MSE versus SNR (dB) using hybrid measurement matrix with $K=8$ and (a) $M=128$, $N=512$; (b) $M=128$, $N=256$.}
	\label{cssMSEmnchange2} 
\end{figure}

In Figs. \ref{csspodMNchange2}(a) and \ref{cssMSEmnchange2}(a), we set the number of measurements $M$ to be half of that in Fig. \ref{css2}. It is observed that the performance of all the algorithms decrease compared with that in Fig. \ref{css2}, while our proposed B-OLS-CSS still approaches to CoSaMP-CSS and MOLS-CSS. In Figs. \ref{csspodMNchange2}(b) and \ref{cssMSEmnchange2}(b), the number of columns $N$ of the measurement matrix is further set to be half of that in Figs. \ref{csspodMNchange2}(a) and \ref{cssMSEmnchange2}(a), which leads to a smaller matrix coherence. In this case, the performance of OMP-CSS, B-OMP-CSS, OLS-CSS and B-OLS-CSS improves significantly. Note that the performance of our proposed B-OLS-CSS is closer to that of the CoSaMP-CSS and MOLS-CSS compared with the results in Figs. \ref{csspodMNchange2}(a) and \ref{cssMSEmnchange2}(a).

\section{Conclusion}
This work is motivated by the challenge in practical CR that the prior information such as the spectrum sparsity and noise variance is usually unavailable for CSS techniques. To address this issue, we propose a B-OLS-CSS algorithm, which works properly even without such prior information. The theoretical analysis demonstrates that the ${\rm SNR}$ required for reliable recovery of our B-OLS-CSS is lower than the existing blind algorithm, leading to theoretical guarantee for the algorithm robustness under the low SNR environments. Simulation results verify that the proposed B-OLS-CSS algorithm can provide comparable performance to the one using sparsity and noise information, and it outperforms the other existing blind CSS techniques in terms of better CSS accuracy.


%

\appendices
\section{Proof of Lemma \ref{lemma3}}
\label{proofoflemma3}
\begin{IEEEproof}
Due to submultiplicativity, we have $||\mathbf{P}_{\mathbf{S}^l}\mathbf{D}_i||_2\leq\rho(\mathbf{D}_{\mathbf{S}^l})\rho((\mathbf{D}_{\mathbf{S}^l}^T\mathbf{D}_{\mathbf{S}^l})^{-1})||\mathbf{D}_{\mathbf{S}^l}\mathbf{D}_i||_2$.
Based on Lemma \ref{lemma1}, we obtain $\rho(\mathbf{D}_{\mathbf{S}^l})
=\sqrt{\lambda_{\max}(\mathbf{D}^T_{\mathbf{S}^l}\mathbf{D}_{\mathbf{S}^l})}\leq\sqrt{1+\sqrt{K/M}+\rho}$.
Then, similar to the proof procedures in \cite{2}, we have $\rho((\mathbf{D}_{\mathbf{S}^l}^T\mathbf{D}_{\mathbf{S}^l})^{-1})
\leq\frac{1}{1-\sqrt{K/M}-\rho}$.
Since $||\mathbf{D}_{\mathbf{S}^l}\mathbf{D}_i||_2\leq\sqrt{K\mu^2}$, $||\mathbf{P}_{\mathbf{S}^l}\mathbf{D}_i||_2\leq\frac{\sqrt{K\mu^2(1+\sqrt{K/M}+\rho)}}{1-\sqrt{K/M}-\rho}$.
Finally, the proof is completed because $||\mathbf{P}^{\bot}_{\mathbf{S}^l}\mathbf{D}_i||_2=\sqrt{1-||\mathbf{P}_{\mathbf{S}^l}\mathbf{D}_i||_2^2}$.
\end{IEEEproof}

\section{Proof of Theorem \ref{theorem4}}
\label{proofoftheorem4}

\begin{IEEEproof}
The proof of Theorem \ref{theorem4} contains three parts: 1) Developing the condition for choosing a correct entry in each iteration; 2) Proving that the OLS algorithm does not stop at the $l$-th iteration ($l<K$); 3) Proving that the OLS algorithm stops after $K$ iterations.

We now prove the first point. According to \cite[Lemma 4]{4} and \cite[Theorem 5]{2}, we obtain that if
\begin{equation}\label{theorem52}
||\mathbf{x}_{0\backslash\mathbf{S}}||_{2}>\frac{2\sqrt{K-l}(2-(K-\mathcal{T})\mu)\omega\mu\sigma\theta}{(2-(K-\mathcal{T})\mu-2K\mathcal{T}\mu)(1-(K-1)\mu)},
\end{equation}
OLS selects a correct atom with the probability ${\rm P}\{||\mathbf{D}^T\mathbf{P}_{\mathbf{S}^l}^{\bot}\mathbf{\epsilon}||_{\infty}\leq\omega\mu\theta\sigma\}$.
Since $\frac{1}{\sigma}||\mathbf{x}_{0\backslash\mathbf{S}}||_{2}=\sqrt{\sum_{q\in0\backslash\mathbf{S}}M\times{\rm SNR}_q}$,
(\ref{theorem52}) becomes ${\rm SNR_{\min}}>\varphi_1$.

Next, based on \cite[Lemma 5.1]{6}, for $l<K$, we obtain
\begin{align}
&\frac{||\mathbf{D}^T\mathbf{r}^l||_\infty}{||\mathbf{r}^l||_2}
\geq\frac{\frac{1}{\sqrt{K-l}}||\mathbf{D}^T_{0\backslash\mathbf{S}}\mathbf{P}^{\bot}_{\mathbf{S}^l}\mathbf{D}_{0\backslash\mathbf{S}}\mathbf{x}_{0\backslash\mathbf{S}}||_2-\omega\mu\theta\sigma}{||\mathbf{P}^{\bot}_{\mathbf{S}^l}\mathbf{D}_{0\backslash\mathbf{S}}\mathbf{x}_{0\backslash\mathbf{S}}||_2+\sqrt{M+2\sqrt{M\log M}}\sigma}\nonumber\\
\geq&\frac{\frac{1-\sqrt{K/M}-\vartheta}{\sqrt{K-l}}||\mathbf{x}_{0\backslash\mathbf{S}}||_2-\omega\mu\theta\sigma}{(1+\sqrt{K/M}+\vartheta)||\mathbf{x}_{0\backslash\mathbf{S}}||_2+\sqrt{M+2\sqrt{M\log M}}\sigma}\label{theorem57}
\end{align}
with the probability ${\rm P}\Big\{\bigcap\limits^{K}_{l=1}||\mathbf{D}^T\mathbf{P}_{\mathbf{S}^l}^{\bot}\mathbf{\epsilon}||_{\infty}\leq\omega\mu\theta\sigma,\zeta_{\min}\geq1-\sqrt{K/M}-\vartheta, \zeta_{\max}\leq1+\sqrt{K/M}+\vartheta,||\epsilon||_2\leq\sqrt{M+2\sqrt{M\log M}}\sigma\Big\}$.
Then, if ${\rm SNR_{\min}}>\varphi_2$,
we have $\frac{||\mathbf{D}^T\mathbf{r}^l||_\infty}{||\mathbf{r}^l||_2}>\omega\mu$. Thus, OLS does not stop at the $l$-th iteration.

Now it remains to prove that OLS stops exactly at the $K$-th iteration.
By using \cite[Lemma 3]{4}, with the probability ${\rm Pr}\{||\mathbf{D}^T\mathbf{P}^{\bot}_{\mathbf{S}^K}\mathbf{\epsilon}||_{\infty}\leq\omega\mu\theta\sigma,||\mathbf{P}^{\bot}_{\mathbf{S}^K}\mathbf{\epsilon}||_2\geq\theta\sigma\}$, we have
\begin{equation}\label{theorem56}
\begin{aligned}
\frac{||\mathbf{D}^T\mathbf{r}^K||_\infty}{||\mathbf{r}^K||_2}=\frac{||\mathbf{D}^T\mathbf{P}_{\mathbf{S}^K}^{\bot}\mathbf{\epsilon}||_{\infty}}{||\mathbf{P}_{\mathbf{S}^K}^{\bot}\mathbf{\epsilon}||_2}\leq\omega\mu,
\end{aligned}
\end{equation}
which means that OLS stops at the $K$-th iteration. Finally, based on the aforementioned analysis and due to $K<\mathcal{C}$, the proof is completed by writing out the above probability.
\end{IEEEproof}
\section*{Acknowledgment}

This work was supported by the National Natural Science Foundation of China (61871050), US National Science Foundation (2003211, 2128596, and 2136202) and Virginia Research Investment Fund (CCI-223996).

\ifCLASSOPTIONcaptionsoff
  \newpage
\fi



%

\bibliographystyle{IEEEtran}
\bibliography{TIT}

\begin{thebibliography}{10}
\providecommand{\url}[1]{#1}
\csname url@samestyle\endcsname
\providecommand{\newblock}{\relax}
\providecommand{\bibinfo}[2]{#2}
\providecommand{\BIBentrySTDinterwordspacing}{\spaceskip=0pt\relax}
\providecommand{\BIBentryALTinterwordstretchfactor}{4}
\providecommand{\BIBentryALTinterwordspacing}{\spaceskip=\fontdimen2\font plus
\BIBentryALTinterwordstretchfactor\fontdimen3\font minus
  \fontdimen4\font\relax}
\providecommand{\BIBforeignlanguage}[2]{{%
\expandafter\ifx\csname l@#1\endcsname\relax
\typeout{** WARNING: IEEEtran.bst: No hyphenation pattern has been}%
\typeout{** loaded for the language `#1'. Using the pattern for}%
\typeout{** the default language instead.}%
\else
\language=\csname l@#1\endcsname
\fi
#2}}
\providecommand{\BIBdecl}{\relax}
\BIBdecl

\bibitem{64}
E.~Hill and H.~Sun, ``Double threshold spectrum sensing methods in
  spectrum-scarce vehicular communications,'' \emph{IEEE Trans. Ind. Inf.},
  vol.~14, no.~9, pp. 4072--4080, Sep. 2018.

\bibitem{4}
S.~Chen, Z.~Cheng, C.~Liu, and F.~Xi, ``A blind stopping condition for
  orthogonal matching pursuit with applications to compressive sensing radar,''
  \emph{Signal Process.}, vol. 165, no. Dec., pp. 331--342, 2019.

\bibitem{56}
J.~Kim \emph{et~al.}, ``Optimal restricted isometry condition of normalized
  sampling matrices for exact sparse recovery with orthogonal least squares,''
  \emph{IEEE Trans. Signal Process.}, vol.~69, pp. 1521--1536, 2021.

\bibitem{55}
Y.~Luo, J.~Dang, and Z.~Song, ``Optimal compressive spectrum sensing based on
  sparsity order estimation in wideband cognitive radios,'' \emph{IEEE Trans.
  Veh. Technol.}, vol.~68, no.~12, pp. 12\,094--12\,106, Dec. 2019.

\bibitem{60}
J.~Yu, X.~Liu, H.~Qi, and Y.~Gao, ``Long-term channel statistic estimation for
  highly-mobile hybrid {mmWave} multi-user {MIMO} systems,'' \emph{IEEE Trans.
  Veh. Technol.}, vol.~69, no.~12, pp. 14\,277--14\,289, Dec. 2020.

\bibitem{2}
L.~Lu, W.~Xu, Y.~Wang, and Z.~Tian, ``Recovery conditions of sparse signals
  using orthogonal least squares-type algorithms,'' \emph{IEEE Trans. Signal
  Process.}, vol.~70, pp. 4727--4741, Sep. 2022.

\bibitem{9}
C.~Soussen, R.~Gribonval, J.~Idier, and C.~Herzet, ``Joint $k$-step analysis of
  orthogonal matching pursuit and orthogonal least squares,'' \emph{IEEE Trans.
  Inf. Theory}, vol.~59, no.~5, pp. 3158--3174, May 2013.

\bibitem{65}
J.~Wen \emph{et~al.}, ``Nearly optimal bounds for orthogonal least squares,''
  \emph{IEEE Trans. Signal Process.}, vol.~65, no.~20, pp. 5347--5356, Oct.
  2017.

\bibitem{1}
T.~T. Cai and L.~Wang, ``Orthogonal matching pursuit for sparse signal recovery
  with noise,'' \emph{IEEE Trans. Inf. Theory}, vol.~57, no.~7, pp. 4680--4688,
  Jul. 2011.

\bibitem{57}
K.~R. Davidson and S.~J. Szarek, ``Local operator theory, random matrices and
  banach spaces,'' \emph{In W.B. Johnson and J. Lindenstrauss, Handbook of the
  geometry of Banach spaces}, pp. 317--366, 2002.

\bibitem{15}
J.~Wang and P.~Li, ``Recovery of sparse signals using multiple orthogonal least
  squares,'' \emph{IEEE Trans. Signal Process.}, vol.~65, no.~8, pp.
  2049--2062, 2017.

\bibitem{6}
T.~T. Cai \emph{et~al.}, ``On recovery of sparse signals via $\ell_1$
  minimization,'' \emph{IEEE Trans. Inf. Theory}, vol.~55, no.~7, pp.
  3388--3397, Jul. 2009.

\end{thebibliography}

%




\end{document}